\documentclass[a4paper]{EslabStyle}
\usepackage{epsfig}

\newcommand{\ms}{M$_{\odot}$}

\begin{document}

\title{The role of Star Formation in the evolution of spiral galaxies}

\author{N. Prantzos\inst{1}}
  \institute{Institut d'Astrophysique de Paris, 98bis Bd. Arago, 75014 Paris}

\maketitle 

\begin{abstract}

 Spiral galaxies offer a unique opportunity to study the role of star
formation in galaxy evolution and to test various theoretical 
star formation schemes. I review some recent relevant work on the evolution
of spiral galaxies. Detailed models are used for the chemical and 
spectrophotometric evolution, with metallicity dependent stellar
yields, tracks and spectra. The models are ``calibrated'' on the
Milky Way disk and generalised to other spirals with some simple
scaling relations, obtained in the framework of Cold Dark Matter
models for galaxy formation. The results compare favourably
to the main observables of present day spirals, provided a crucial
assumption is made: {\it massive disks form their stars earlier
than low mass ones}. It is not clear whether this picture is
compatible with currently popular hierarchical models of galaxy
evolution. The resulting abundance gradients are found to be
anticorrelated to the disk scalelength, support radially
dependent star formation efficiencies and point to a kind
of ``homologuous evolution'' for spirals


\end{abstract}

\section{Introduction}

Among the three main ingredients required for galactic evolution
studies (stellar evolution data, stellar initial mass function
and star formation rate), the star formation rate (SFR)
is the most important and the less well understood. Despite more than
30 years of observational and theoretical work, the Schmidt law still 
remains popular among theoreticians and compatible with most available
observations.
Up to the mid-90ies, observations revealed features concerning the
present status of nearby galaxies, from which their past history
had to be derived. Developments in the past few years revealed
features of the ``global SFR'' of the Universe and opened
perspectives for seing directly the past SFR history of the 
various galaxy types.

Among the various galaxy types, the disks of spirals are certainly the ones
offering the best opportunities to study star formation, for several
reasons: i) with respect to ellipticals, spirals offer more constraints
(amounts of gas and SFR, metal abundances in the gas); ii) with respect
to irregulars, spirals present gradients of various quantities (metal
abundances, colours, gas and SFR profiles), which may constrain various
theoretical schemes for the local SFR; iii) last, but not least, our own
Milky Way is a typical spiral, rather well understood.

In this paper, I will summarize some recent work made on the role
of star formation in the evolution of spiral galaxies. In Sec. 2, a brief
overview is presented of the current status of SFR observations in spirals,
based mostly on the work of Kennicutt (1998); in particular, it is pointed
out that a radial dependence of the star formation efficiency is compatible
both with observations of the Milky Way and with theoretical ideas.
In Sec. 3 I present a model developed recently for the study of the
chemical and spectrophotomeric evolution of disk galaxies. The model
utilises metallicity dependent stellar evolution data (yields, tracks, 
spectra), a radially varying SFR ``calibrated'' on the Milky Way
and describes galactic disks as a
2-parameter family (the two parameters being: maximal rotational velocity 
$V_C$ and spin parameter $\lambda$).  
The model compares favourably to all currently available observations of
nearby spirals, provided a crucial assumption is made:
{\it massive disks form their stars earlier than low mass ones}.
This is supported by the finding that low mass disks have
today larger gas fractions than their high mass counterparts, despite
the fact that the star formation efficiency seems to be independent
of galactic mass.

\section{Star formation in disk galaxies}

A nice account of the various diagnostic methods used to derive
the SFR in galaxies is given in the recent review of Kennicutt (1998),
on which most of the material of this section is based.

Integrated colours and spectra were used in the past to estimate
the SFR, through population synthesis models. This method is still
applied to multi-colour observations of faint galaxies, but has
been supplanted by direct tracers of SFR: measurements of integrated light 
in the UV and the FIR, or of nebular recombination line intensities.
Each one of the direct tracers presents its own advantages and drawbacks,
and an appropriate calibration is not always easy. 

\begin{figure}[ht]
  \begin{center}
    \epsfig{file=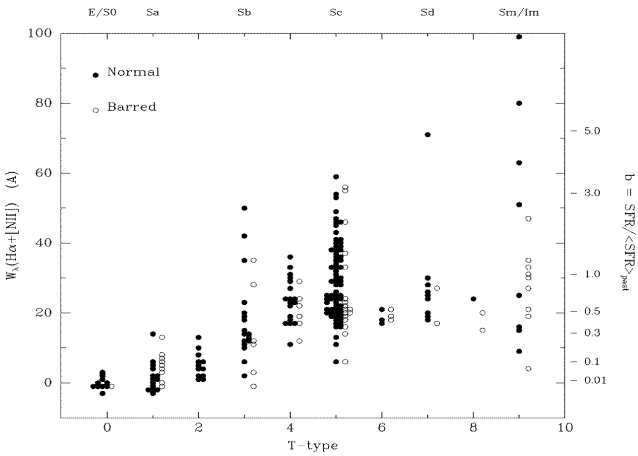, width=6cm, angle=-90}
  \end{center}
\caption{
Distribution of integrated H + [NII] emission-line
equivalent widths (a measure of star formation activity normalised to
galaxy's mass)
for a large sample of nearby spiral galaxies,
subdivided by Hubble type and bar morphology. The right axis
scale shows corresponding values of the stellar birthrate
parameter b, which is the ratio of the present SFR to that
averaged over the past (from Kennicutt 1998).
}
\end{figure}

The UV continuum
(optimal range: 1250-2500 \AA) traces directly the emissivity of a 
young stellar population and can be applied to star-forming galaxies over
a wide range of redshifts. It is, however, very sensitive to extinction,
and the relevant corrections (which may reach several magnitudes) are
difficult to calibrate. This is also true for the diagnostics based on
recombination lines; these lines re-emit effectively the
integrated stellar luminosity of galaxies shortward of the Lyman
limit; since ionizing flux is produced almost exclusively by stars
with M$>$10 \ms, the derived SFR is very sensitive to the slope
of the upper mass part of the adopted IMF. In fact, sensitivity to the
IMF is a common feature of all SFR diagnostic methods.
In the case of the FIR continuum (emitted by interstellar dust that has
been heated by absorption of stellar light) the situation is
complicated by the fact that dust may either surround young star-forming 
regions
or be heated by the general interstellar radiation field; in the former case
FIR really measures the SFR, but the situation is less clear in the latter.
FIR seems then to provide an excellent measure of the SFR in dusty 
circumnuclear starbursts, but its application to normal disk galaxies is not
straightforward.

These techniques have been used to derive the SFR in hundreds of nearby
galaxies, thus revealing several interesting trends of star formation along the
Hubble sequence. In Fig. 1 is plotted the equivalent width (EW) of the
H$\alpha$+[NII] lines (defined as the emission line luminosity devided by the
adjacent continuum flux), which is a measure of the relative SFR, i.e. the
SFR normalised by unit stellar luminosity (in the red). 
There is a strong dispersion at each Hubble type (by a factor of $\sim$10), 
but also a trend of increase of the mean value, by a factor of $\sim$20
between types Sa and Sc.

\begin{figure}[ht]
  \begin{center}
    \epsfig{file=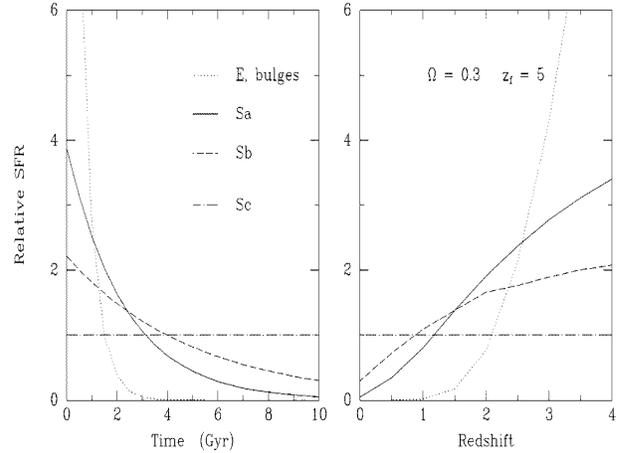, height=8cm,width=6cm, angle=-90}
  \end{center}
\caption{ Schematic illustration of the evolution of the stellar
birthrate for different Hubble types. The left panel shows the
evolution of the relative SFR with time, following Sandage
(1986). The curves for spiral galaxies are exponentially
declining SFRs that fit the mean values of the birthrate
parameter b measured by Kennicutt et al (1994). The curve for
elliptical galaxies and bulges corresponds to an
e-folding time of 0.5 Gyr, and is shown for comparative purposes only. The
right panel shows the corresponding evolution in SFR with
redshift, for an assumed cosmological density parameter  $\Omega$= 0.3 
and an assumed formation redshift z = 5 (from Kennicut 1998). 
}
\end{figure}

This trend should reflect the underlying past star formation histories
of disks, which can be parametrised by $b$, the ratio of current SFR to 
the past SFR averaged over the age of the disk (i.e. Scalo 1986).
Parameter $b$ (derived on the basis of population synthesis models)
appears on the right axis scale of Fig. 1. Typical late type disks have formed
stars at roughly constant rates ($b\sim$1) while early type spirals formed
their stars quite early ($b\sim$0.01-0.1). 

A schematic illustration of the corresponding star formation histories is 
presented in Fig. 2. On the left panel are shown the histories of typical
ellipticals (and bulges of spirals) and of the disks of Sa, Sb and Sc 
galaxies. Obviously, in this simple picture, the Hubble sequence is uniquely
determined by the caracteristic time scale of star formation.

This simple description of the Hubble sequence fails to answer two important
(and probably related) questions: i) what determines the caracteristic time
scales for star formation? and ii) what is the role (if any) of a galaxy's mass
in this picture? Gavazzi et al. (1996) have found an anti-correlation 
between the SFR per unit mass and the galaxy luminosity. Part of this trend may
reflect the same dependence of SFR on Hubble type shown in Fig. 1, but it may
also be that these trends are fundamentally related to the galactic mass.
We shall come to this point in the next section.

\begin{figure}[ht]
  \begin{center}
    \epsfig{file=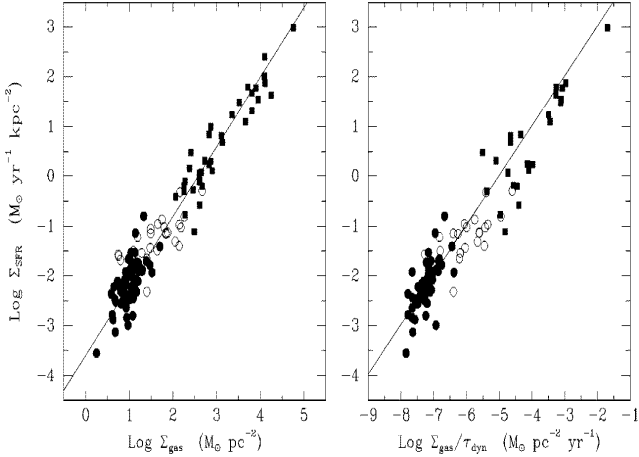, height=8.4cm,width=6cm, angle=-90}
  \end{center}
\caption{Left: The global Schmidt law in galaxies. Solid points
denote normal spirals, squares denote 
circumnuclear starbursts. The open circles show
the SFRs and gas densities of the central regions of the normal
disks. Right: The same SFR data are plotted against the ratio
of the gas density to the average orbital time in the disk 
(from Kennicutt 1998). 
}
\end{figure}

Although it is not yet clear what is the main factor affecting the SFR, it 
seems that on large scales there is a correlation between the SFR and the
gaseous content of galaxies and starbursts. When the {\it average} SFR 
density $\langle\Psi\rangle$ is plotted vs. the average gas surface density 
$\langle\Sigma_G\rangle$,
a Schmidt-type law is recovered:  
$\langle\Psi\rangle \propto \langle\Sigma_G\rangle^N$, 
with $N\sim$1.5 (Fig. 3 left panel).
Such a law is expected if the SFR is proportional to $\rho/\tau_{FF}$
(where $\rho$ is the mass density and $\tau_{FF}\propto\rho^{-1/2}$ is the
free-fall time scale) and provided the disk has a constant scale
height (such as $\langle\Sigma_G\rangle\propto\rho$).
However, Kennicut (1998) finds that the data in Fig. 3 can be equally well
fit by $\langle\Psi\rangle \propto \langle\Sigma_G\rangle \Omega$, 
where $\Omega$ is the frequency
of rotation of the disk measured at half the outer radius $R_{OUT}$.
As noticed by Kennicutt (1998), these different parametrisations lead
to different interpretations for the {\it average efficiency} 
$\epsilon = \langle\Psi\rangle/\langle\Sigma_G\rangle $ 
of star  formation. In the former case one has $\epsilon \propto 
\langle\Sigma_G\rangle^{(N-1)}$, i.e. the efficiency is determined by the
average gas surface density; in the latter case we have $\epsilon \propto
 \Omega$, i.e. $\epsilon$ is determined by how much mass contributes
to the gravitational potential inside  $R_{OUT}$/2.

These findings  are extremely useful for one-zone models
of galaxy evolution (treating the galaxy as a  whole). However,
they are of limited help for multi-zone models of disk evolution, which require
prescriptions for the {\it local} SFR $\Psi(R)$, 
not the average one. If the local
SFR is non-linear w.r.t. the gas surface density, the resulting average
SFR over the whole disk will have no simple relation to the local one.
Although $\langle\Psi\rangle$ can be derived from $\Psi(R)$, the inverse does not hold.

\begin{figure}[ht]
  \begin{center}
    \epsfig{file=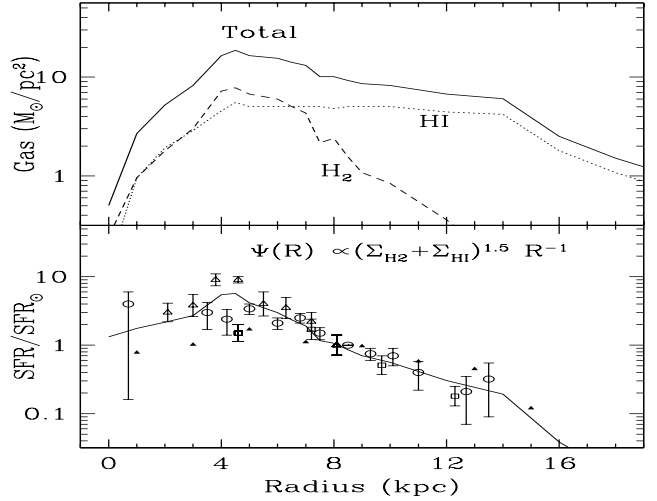, height=7cm, width=8.9cm}
  \end{center}
\caption{
{\it Upper panel:} Current surface density profiles of molecular (H$_2$) and
atomic (HI) hydrogen in the Milky Way, as a function of galactocentric radius
and total gas surface density (the sum of the two, increased
by 40\% to account for helium).
{\it Lower panel}: Corresponding theoretical current SFR ({\it solid} curve)
and comparison to observational estimates of the
current SFR profile.
The SFR profiles are normalised to their value at $R_S$=8 kpc
(from Boissier and Prantzos 1999).
\label{Fig4}}
\end{figure}

A radial dependence of
the star formation efficiency in disk galaxies is required in order
to reproduce the observed abundance and colour gradients (see Sec. 4), 
and such a dependence
has indeed been proposed   on the basis of various
instability criteria for gaseous disks  (e.g. Talbot and Arnett 1975;
Onihishi 1975; Wyse and Silk 1989; Dopita and Ryder 1994). 
It should be noted that
star formation theories exist and may be tested mainly for disk
galaxies, not for e.g. ellipticals or irregulars. 
A star formation rate  explicitly dependent on
galactocentric radius has been suggested by Onihishi (1975). It is
based on the idea that stars are formed
in spiral galaxies when the interstellar medium with angular frequency
$\Omega(R)$ is periodically compressed by the passage of the spiral
pattern, having a frequency $\Omega_P$=const.$<< \Omega(R)$. This leads
to SFR $\propto \Omega(R)-\Omega_P \propto \Omega(R)$ and,
for disks with flat rotation curves, to SFR $\propto$ R$^{-1}$
(Wyse and Silk 1989). In recent works  (Prantzos and Silk 1998, 
Boissier and Prantzos 1999) we used for the Milky Way disk
a local SFR of the form:
\begin{equation}
 \Psi(R) \ = \ 0.1 \ \Sigma_G(R)^{1.5} \ (R/R_S)^{-1} 
\ {\rm M_{\odot} pc^{-2} Gyr^{-1}} 
\end{equation}
where $R_S$=8. kpc is the distance of the Sun to the Galactic centre.
Such a radial dependence of the SFR is also compatible with observational
evidence, as can be seen in Fig. 4, displaying
the current gas surface density profile in the Milky Way disk
(upper panel) and the corresponding SFR
(lower pannel); comparison of this theoretical SFR to observations
(lower panel in Fig. 4) shows a fairly good agreement. 
It has been shown that this form of the SFR can
account for the observed gradients of gas fraction, SFR and chemical abundance
profiles in the Milky Way (Boissier and Prantzos 1999).

\section{A model for disk evolution}

In our models
the galactic disk is simulated as an ensemble of concentric, independently
evolving rings, gradually built up by infall of primordial composition. The
chemical evolution of each zone is followed by solving the appropriate
set of integro-differential equations (Tinsley 1980), 
without the Instantaneous Recycling
Approximation. Stellar yields are from Woosley and Weaver (1995) 
for massive stars
and Renzini and Voli (1981) for intermediate mass stars. Fe producing SNIa are
included, their rate being calculated with the prescription of Matteucci and
Greggio (1986). The adopted stellar IMF
is a multi-slope power-law between 0.1 \ms \ and 100 \ms \ from the work of
Kroupa et al. (1993).

The spectrophotometric evolution is followed in a self-consistent way, i.e.
with the  SFR $\Psi(t)$ and metallicity $Z(t)$ of each zone determined 
by the chemical evolution,
and the same IMF. The stellar lifetimes, evolutionary tracks and spectra are
metallicity dependent; the first two are from the Geneva library 
(Schaller  et al. 1992, Charbonnel et al. 1996) and the latter from 
Lejeune et al. (1997). Dust absorption is
included according to the prescriptions of  
Guiderdoni et al. (1998) and assuming a ``sandwich''
configuration for the stars and dust layers (Calzetti et al. 1994).

\subsection{Scaling relations and results}

In order to extend the Milky Way model to other disk galaxies we adopt  the 
``scaling properties'' derived by Mo, Mao and White (1998, hereafter MMW98) 
in the framework of the Cold Dark Matter (CDM) scenario for galaxy formation. 
According to this scenario, primordial density fluctuations give rise to 
haloes of non-baryonic dark
matter of mass $M$, within which baryonic gas condenses later and forms disks
of maximum circular velocity $V_C$. 
It turns out that disk profiles can be expressed in terms of only two 
parameters: rotational velocity $V_C$ 
(measuring the mass of the halo and, by assuming a 
constant halo/disk mass ratio, also the mass of the disk) and spin 
parameter $\lambda$ (measuring the specific angular momentum of the halo).
In fact, a third parameter, the redshift of galaxy formation (depending
on galaxy's mass) is playing a key role in all hierarchical models of galaxy
formation. However, since the term ``time of galaxy formation'' is not
well defined (is it the time that the first generation of stars form? or
the time that some fraction of the stars, e.g. 50\%, form?) we prefer
to ignore it and assume that 
{\it all disks  start forming their stars at the
same time, but not at the same rate}. In that case
the profile of a given disk 
(caracterised by the central surface density $\Sigma_0$ and the disk
scalelength $R_d$)
can  be expressed in terms of the one of our
Galaxy (the parameters of which are designated hereafter by index G):

\begin{figure}[h]
  \begin{center}
    \epsfig{file=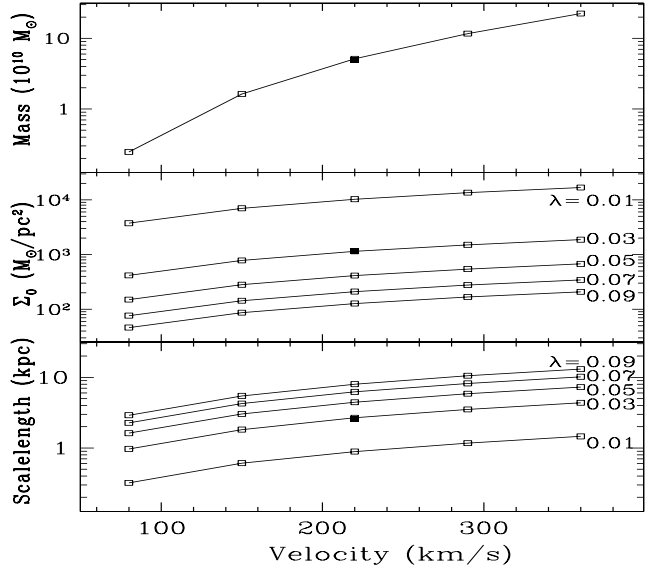, height=8cm, width=8.9cm}
  \end{center}
\caption{Main properties of our model disks. From top to
bottom: mass, central surface density and scalelength, respectively.
They are plotted as a function of circular velocity $V_C$ and 
parametrised with the spin parameter $\lambda$. 
Disk mass depends only on $V_C$. Filled symbols correspond
to the Milky Way model, used for the scaling of all other disks
(from Boissier and Prantzos 2000).
\label{fig1}}
\end{figure}

\begin{equation}
\frac{R_d}{R_{dG}}  \  = \  \frac{\lambda}{\lambda_G}  \frac{V_C}{V_{CG}}
\ \ \ \  {\rm and} \ \ \ \  
\frac{\Sigma_0}{\Sigma_{0G}}  \  =  \left(\frac{\lambda}{\lambda_G}\right)^{-2}
 \frac{V_C}{V_{CG}}
\end{equation}
where we have adopted $\lambda_G$=0.03. The absolute value
of $\lambda_G$ is of little importance as far as it is close
to the peak of he corresponding distribution function (see
next paragraph), since our disk profiles and 
results depend only  on the ratio $\lambda/\lambda_G$.

Eqs. 2  allow  to describe the mass profile of a galactic disk
in terms of the one of our Galaxy and of two parameters: $V_C$ and $\lambda$.
The range of observed values for the former parameter
is 80-360 km/s, whereas for the latter
numerical simulations give values in the 0.01-0.1 range, the distribution
peaking around $\lambda\sim0.04$ (MMW98).  
Although it is not clear yet whether
$V_C$ and $\lambda$ are independent quantities, we treat them here as such
and construct a grid of 25 models caracterised by $V_C$ = 80, 150, 220, 290, 360 km/s
and $\lambda$ = 0.01, 0.03, 0.05, 0.07, 0.09, respectively.
[{\it Notice:}  if $\lambda_G$=0.06 is adopted for the Milky Way, our 
model results would be the same, but they would correspond to  values of $\lambda$
twice as large, i.e. 0.02, 0.06, 0.10, 0.14 and 0.18, respectively]. 
Increasing  values of $V_C$ correspond to more massive disks and
increasing values of $\lambda$ to more extended disks (Fig. 5).

\begin{figure*}[ht]
  \begin{center}
    \epsfig{file=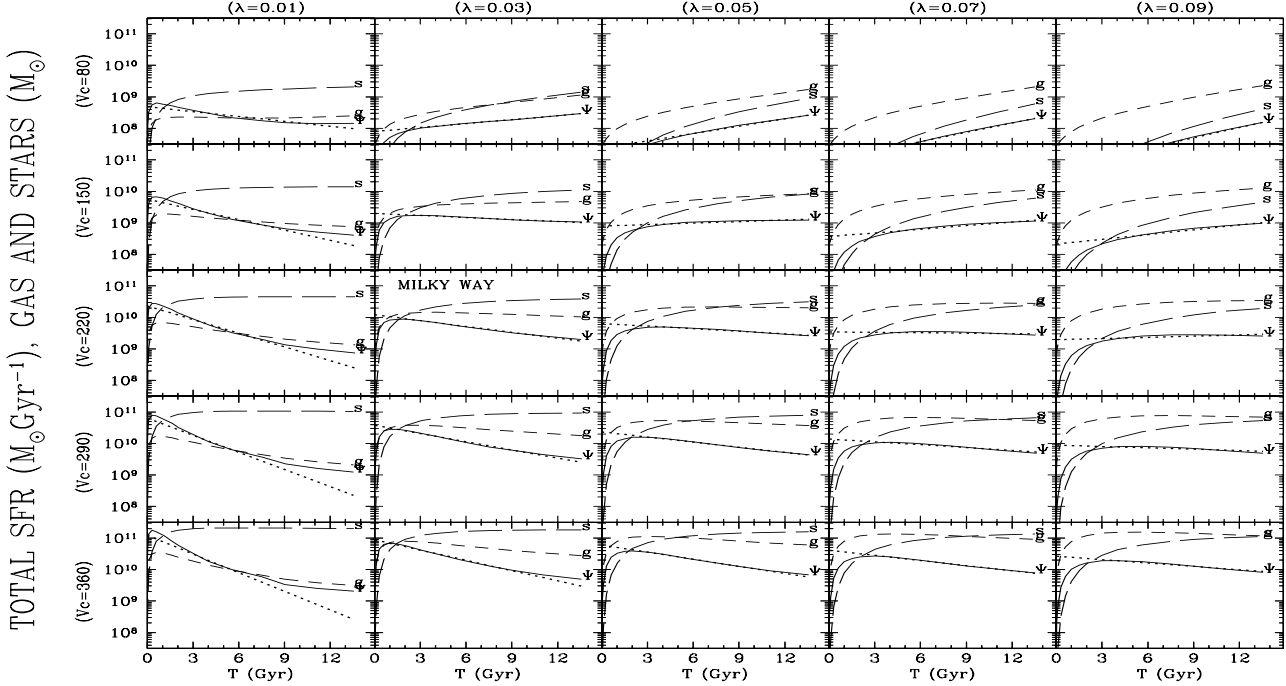,height=18cm, width=10cm,angle=-90}
  \end{center}
\caption{ Histories of
total SFR ($\bf \Psi$, {\it solide curve}), gas mass ($\bf g$, 
{\it short dashed curve})
 and stellar mass ($\bf s$, {\it long dashed curve}) for our disk
models. Each column
corresponds to a value of $\lambda$ and each row to a value of $V_C$.
The {\it dotted} curve is an exponential fit to the SFR 
(from Boissier and Prantzos  2000).
\label{fig1}}
\end{figure*}

We calculate the corresponding rotational velocity curves $V(R)$ of our model disks
assuming contributions from the disk and an isothermal dark halo. The local SFR is then
calculated self-consistently as:
\begin{equation}
 \Psi(t,R) \ = \  \alpha \  \Sigma_g(t,R)^{1.5} \ V(R) \ R^{-1}
\end{equation}
It can be seen that the corresponding star formation efficiency at the caracteristic
radius $R_d$ does not depend on $V_C$ (for a given $\lambda$). In other terms, {\it the
adopted SFR law is important only for the resulting gradients, but not for the
global properties of galaxies}. In our models, the timescale for star formation is
affected by the adopted infall timescale, {\it adjusted as to form massive disks earlier
than less massive ones}; this most important ingredient is discussed in Sec. 3.2 and 4.

The evolution of the SFR and of the gaseous and stellar masses in our 25 disk models
is shown in Fig. 6. This evolution  results from the 
adopted prescriptions for infall and SFR.
Values of SFR range from $\sim$100 $M_{\odot}$/yr in the early phases of massive disks
to $\sim$0.1 $M_{\odot}$/yr for the lowest mass galaxies.
The resulting SFR history is particularly interesting
when compared to simple models of photometric evolution, that are usually 
applied in studies
of galaxy evolution and of their cosmological implications. Such models (one-zone,
no chemical evolution considered in general) adopt exponentially declining
SFR with different caracteristic timescales for each galaxy morphological type 
(e.g. Bruzual \& Charlot 1993, Fioc and Rocca-Volmerange 1997).

The fact that in some cases the SFR increases in time leads us to define the time-scale
$\tau_{\bf *}$,  required for forming the first half of the stars of a given galaxy.
This time scale for star formation appears in Fig. 7, plotted as a
function of $\lambda$ (left) and $V_C$ (right). It can be seen that $\tau_{\bf *}$
is a monotonically increasing function of $\lambda$ and a decreasing function of $V_C$.
For the Milky Way ($V_{CG}$=220 km/s, $\lambda_G$=0.03) we find $\tau_{\bf *}\sim$3.5
Gyr. For galaxies less massive than the Milky Way and $\lambda>0.06$, it  takes more than
half the age of the Universe to      form the first half of their stars.

\begin{figure}[h]
  \begin{center}
    \epsfig{file=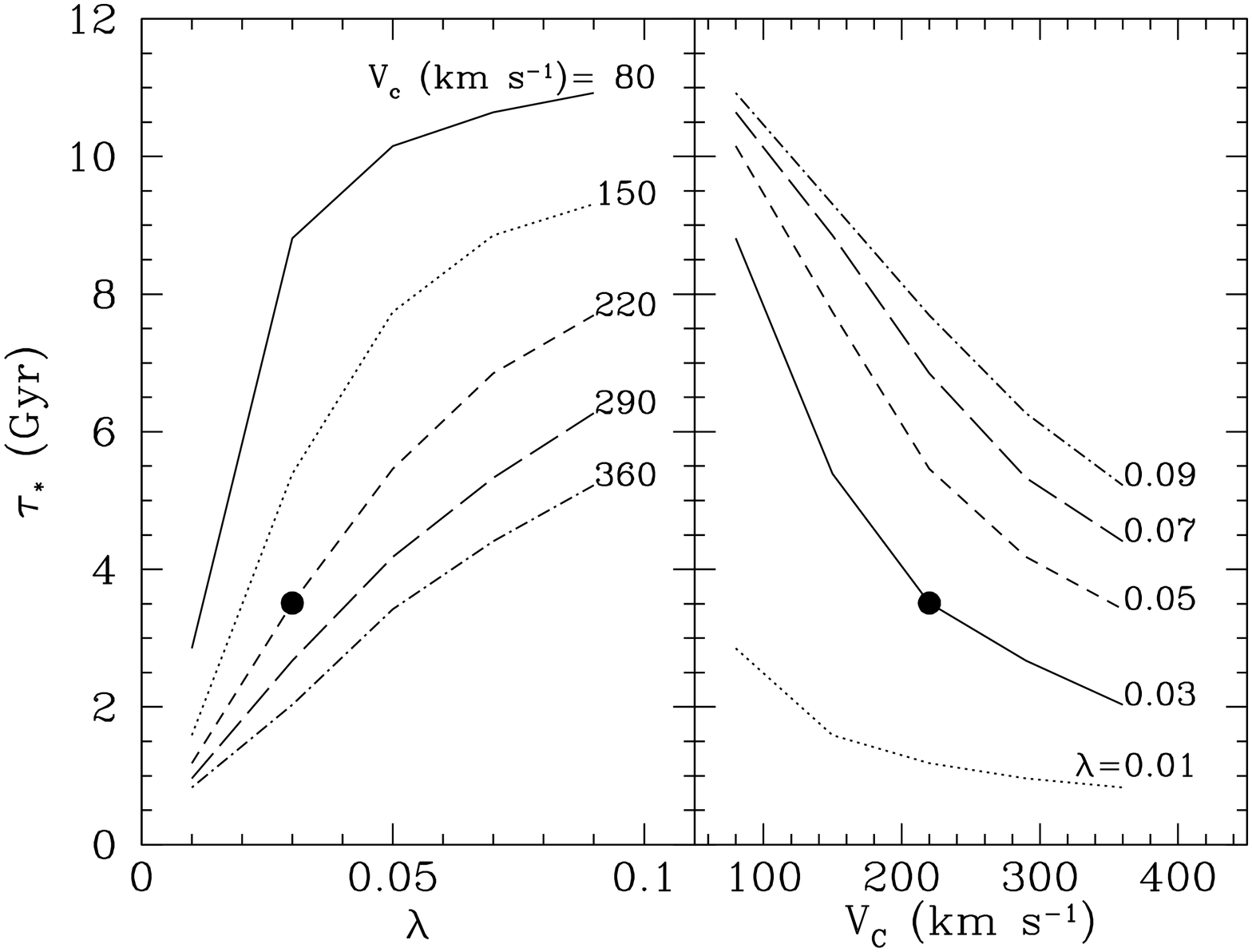, height=8.9cm,width=5.8cm, angle=-90}
  \end{center}
\caption{Caracteristic time-scales for star formation, i.e.
for forming the first half of the stars of a given galaxy, 
resulting from our models. They
are plotted as a function of  the spin parameter $\lambda$ ({\it left}) and of
circular velocity $V_C$ ({\it right}).
{\it Filled points} correspond to the Milky Way
(from Boissier and Prantzos  2000).
}
\end{figure}

\begin{figure*}[ht]
  \begin{center}
    \epsfig{file=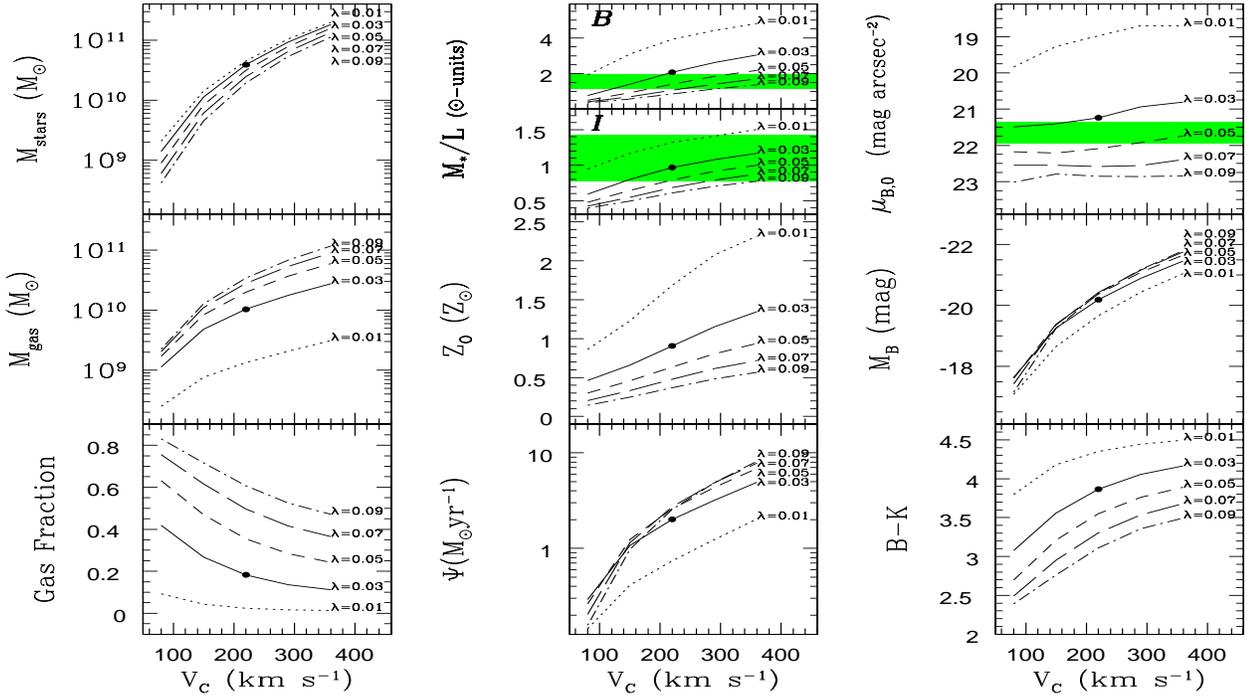, height=18cm, width=10.cm, angle=-90}
  \end{center}
\caption{Integrated properties of our models at an age T=13.5 Gyr.
Mass of stars ({\it top left}),
mass of gas   ({\it middle left}),
gas  fraction ({\it bottom left}),
M/L values in B- and I- bands ({\it top center}), average oxygen abundance, 
i.e. total mass of oxygen divided by total mass of gas
in the disk ({\it middle center}),
star formation rate ({\it bottom center}),
central surface brightness $\mu_{B0}$   ({\it top right}),
total B-magnitude        ({\it middle right}),
total B-K     ({\it bottom right}). All results are plotted as a function of
the rotational velocity $V_C$ and parametrised with the values of the spin
parameter $\lambda$. The {\it filled symbols} on the $\lambda$=0.03 curves
and for $V_C$=220 km/s in all panels  correspond to Milky Way values. 
The {\it grey bands} in the $M/L$ and $\mu_{B0}$ panels indicate
the corresponding ranges of observed values
(from Boissier and Prantzos  2000).
\label{fig1}}
\end{figure*}

Assuming  that all our model disks started forming their stars $\sim$13  Gyr ago,
we present in Fig. 8 their integrated properties today (at redshift $z$=0).
Several interesting results can be pointed out:

1) For a given $\lambda$, the stellar and gaseous masses are monotonically
increasing functions of $V_C$. For a given $V_C$ disks may have considerably
different gaseous and stellar contents (by factors 30 and 5, respectively),
depending on how ``compact'' they are, i.e. on their $\lambda$ value.
The more compact galaxies have always a smaller gaseous content.
This is well illustrated by the behaviour of the gas fraction 
$\sigma_g$({\it bottom left} of Fig. 8). 

2) The current SFR $\Psi_0$ is found to be mostly a function
of $V_C$ alone, despite the fact that the amount of 
gas varies by a factor of $\sim$2-3 for
a given $V_C$. This is due to the fact that more extended disks (higher $\lambda$)
are less efficient in forming stars, because of the $1/R_d$ factor (for a given $V_C$).
This smaller efficiency compensates for the larger gas mass 
available in those galaxies, 
and a unique value is obtained for $\Psi_0$. This immediately implies that little
scatter is to be expected in the B-magnitude vs. $V_C$ relation, and  this is indeed
the case as can be seen in Fig. 8 ({\it middle right panel}). 
For that reason we find litle scatter in the corresponding Tully-Fisher relation,
which is in fair agreement with available data in various wavelength bands
(as discussed in detail in Boissier and Prantzos 2000).

3) The mass to light ratio $\Upsilon=M/L$ is a quantity often used
to probe the age and/or the IMF of the stellar population and the
amount of dark matter, but also to convert results of dynamical models
(i.e. mass) into observed quantities (i.e. light in various wavelength
bands). Our models ({\it upper middle } of Fig. 8)
lead to fairly acceptable values of  $\Upsilon_B$ and $\Upsilon_I$;
these values are confined to a relatively narrow range and  a systematic
(albeit weak) trend is obtained with galaxy's mass, that should not be neglected
in detailed models.

\begin{figure}[h]
  \begin{center}
    \epsfig{file=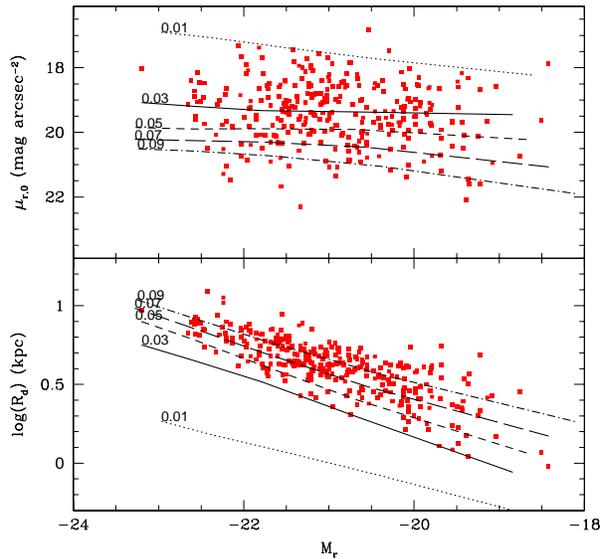, height=8cm, width=8.1cm}
  \end{center}
\caption{Central surface brightness ({\it upper panel})
and disk size $R_d$ ({\it lower panel}) in the r-band, as a function
of the galaxy's magnitude $M_r$. Our results are parametrised by the
values of the spin parameter $\lambda$; $\lambda$=0.01 models produce
unrealistically small  disks with very high surface brightness and 
correspond rather to galactic bulges 
(from Boissier and Prantzos 2000).
}
\end{figure}

4) Most of our model disks have a roughly constant current 
central surface brightness
$\mu_{B0}\sim$21.5 mag arcsec$^{-2}$. This is quite an encouraging result, in view
of the observed constancy of the central surface brightness in disks, around the
``Freeman value'' $\mu_{B0}$=21.7$\pm$0.3 mag arcsec$^{-2}$ (Freeman 1970); this
value is indicated by the grey band in Fig. 8 ({\it top right panel}).
Finally, models with $\lambda$=0.07 and 0.09 result in low surface
brightness (LSB) galaxies today.

\subsection{Comparison to observations}

As explained in Sec. 3.1, our model disks are described by the central surface density
$\Sigma_0$ and scalelength $R_d$. These are related to the ``fundamental''
parameters $V_C$ and $\lambda$ (fundamental because they refer to the dark
matter haloes) through the scaling relations (2), which allow
to use the Milky Way as a ``calibrator''.

In Fig. 9 we present the r-band central surface brightness (upper panel)
and size (lower panel) of our model disks as a function of the galaxy's
luminosity and we compare them to the data of Courteau and Rix (1999).
It can be seen that:

i) The central surface brightness in our models compares fairly well to
the data. For a given value of $\lambda$, $\mu_{r0}$ is quasi-independent on
luminosity (or $V_C$), as explained in Sec. 3.1. The obtained values depend
on $\lambda$, the more compact disks being brighter. All the observed
values are lower than our results of $\lambda$=0.01 models. Higher $\lambda$ values
bracket reasonably well the data.

\begin{figure}[ht]
  \begin{center}
\vspace{-0.5cm}
    \epsfig{file=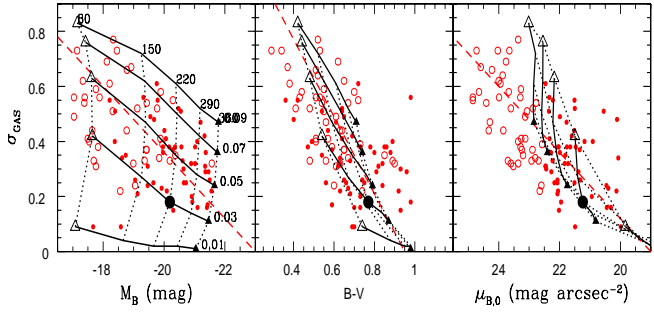, height=9.4cm, width=9cm, angle=-90}
  \end{center}
\vspace{-4.1cm}
\caption{ Gas fraction vs. B-magnitude ({\it left panel}),
B-V colour ({\it middle panel}) and central surface brightness
$\mu_{B0}$ ({\it right panel}).
The grid of models represents 5 values for the spin parameter $\lambda$
({\it solid curves}) and 5 values for the circular velocity
({\it dashed curves}), as indicated on the left panel.
Values of $V_C$ run from 80 km/s ({\it open triangle}) to 360 km/s
({\it filled triangle}).
Results of models are given at a galactic age of 13.5 Gyr.
Data in all panels are from McGaugh and DeBlok (1997).
{\it Filled symbols} correspond to high surface brightness galaxies and
{\it open symbols} to low surface brightness galaxies.
Our grid of models clearly does not fit the lowest surface brightness
galaxies (i.e. below $\mu_{BO}$=23 mag arcsec$^{-2}$), which probably
require values of $\lambda >$0.1.
In all three panels, the {\it dashed} diagonal line is a fit to the
data given in McGaugh and DeBlok (1997): $\sigma_g$ = 0.12 (M$_B$+23) =
-1.4 [(B-V)-0.95] = 0.12 ($\mu_{BO}$-19) (from Boissier and Prantzos 2000).
}
\end{figure}

ii) The size of our model disks decreases  with decreasing luminosity, with
a slope which matches again the observations fairly well. Our disk sizes range 
from $R_d\sim$6-10 kpc for the most luminous galaxies to 1-2 kpc for the less
luminous ones. The $\lambda$=0.01 curve clearly lies too low w.r.t. all data
points, showing once more that this $\lambda$ value
 does not produce realistic disks; it rather produces bulges or elliptical galaxies.

 In Fig. 10 we compare our results to the data of McGaugh and deBlok (1997),
concerning gas fraction vs. various photometric quantities.
The following points can be made:

1) We find indeed a trend between gas fraction and absolute magnitude: smaller
and less luminous galaxies are in general more gas rich, as can also be seen
in Fig. 8 ({\it bottom left} panel). Our grid of models covers reasonnably 
well the range of observed values, except for the lowest luminosity galaxies 
(see also points 2 and 3 below). In particular, the slope of the
$\sigma_g$ vs. M$_B$ relation of our models is similar to the one in the fit
to the data given by McGaugh and deBlok (1997).

2) We find that the observed $\sigma_g$ vs. B-V relation can also be
explained in terms of galaxy's mass, i.e. more massive galaxies are
``redder'' and have smaller gas fractions. Indeed, 
the slope of the $\sigma_g$ vs. B-V relation (-1.4) is again well reproduced
by our models, and the $\lambda$=0.05 models lead to results that match close
the observations. However, although our grid of models reproduces well the 
observed dispersion in $\sigma_g$, it does not cover the full range of B-V
values. The lowest B-V values correspond to gas rich galaxies ($\sigma_g\sim$
0.5-0.8) and could be explained by models with higher $\lambda$ values than
those studies here (i.e. by very Low Surface Brightness galaxies).

\begin{figure}[ht]
  \begin{center}
    \epsfig{file=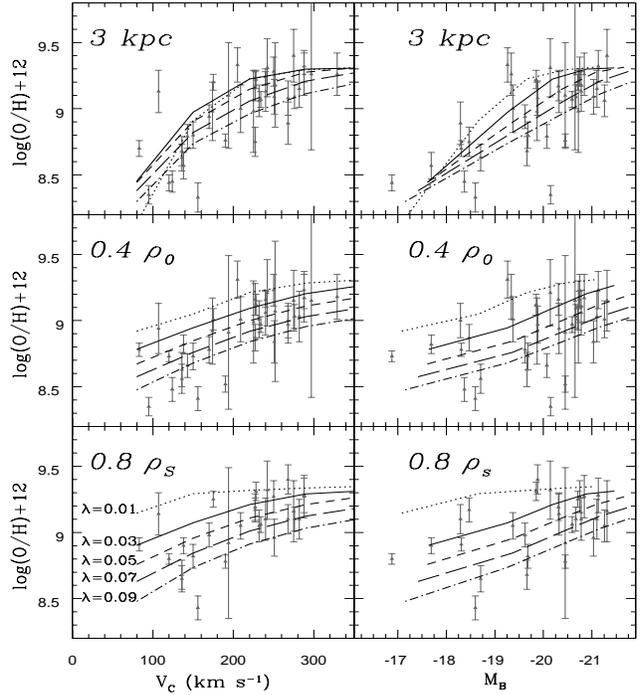, height=10cm, width=8.9cm}
  \end{center}
\caption{Oxygen abundances at various galactocentric distances:
at 3 kpc ({\it upper panels}), 0.4 $\rho_o$ ({\it middle panels}),
and 0.8 $\rho_S$ ({\it lower panels}). They are plotted as a function
of $V_C$ ({\it left}) or, equivalently $M_B$ ({\it right}).
Our results ({\it curves}) are parametrised by the corresponding
 $\lambda$ values.$\rho_0$ is the optical radius and $\rho_S$(=$R_d$
in the rest of the paper) is the disk scalelength.
 Data are from Zaritsky et al. (1994) (from Boissier and Prantzos 2000).
}
\end{figure}

3) The situation is radically different for the $\sigma_g$ vs. $\mu_{B0}$ 
relation. As already seen in Fig. 8 ({\it top right} panel) we find no
correlation in our models between  $\mu_{B0}$ and $V_C$, but we do find
one between $\mu_{B0}$ and $\lambda$: more compact disks (lower $\lambda$)
have higher central surface brightness. It is this latter property that allows 
our models to reproduce, at least partially, the observed $\sigma_g$ vs. 
$\mu_{B0}$ relation ({\it right panel} of Fig. 10).

Another probe of the evolutionary status of a galactic disk is its metallicity.
Both its absolute value and its radial profile can be used as a diagnostic
tool to distinguish between different models of galactic evolution.

In Fig. 11 we plot our results at the same caracteristic radii as Zaritsky
et al. (1994), i.e. at 3 kpc, at 0.4 $\rho_0$ and at 0.8 $R_d$ (from {\it top} to
{\it bottom}, respectively) as a function of circular velocity $V_C$ ({\it
left panels}) and of M$_B$ ({\it right panels}).

\begin{figure}[ht]
  \begin{center}
    \epsfig{file=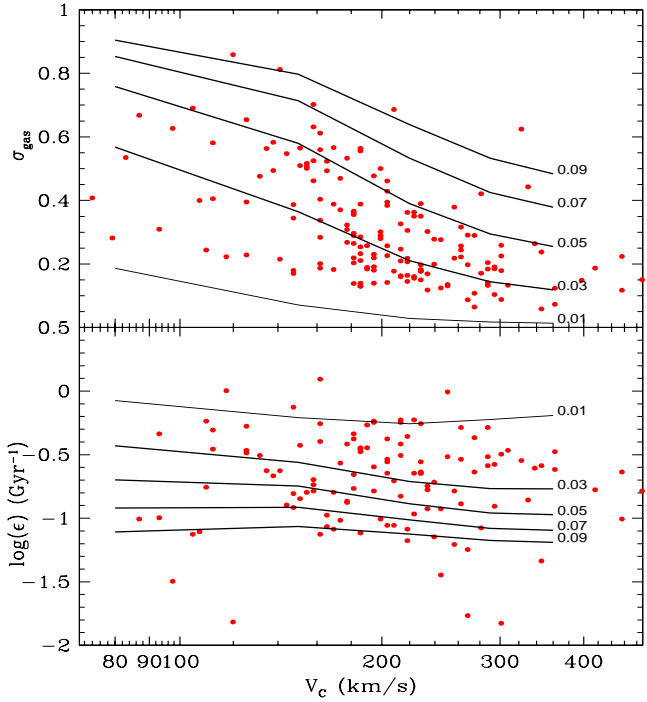, height=10cm, width=8.9cm}
  \end{center}
\caption{ Upper panel : Gas fraction of spiral galaxies as a function of their
circular velocity. Lower panel: Efficiency of star formation (star formation rate
divided by gas mass) for those same galaxies.
Model results are parametrised by teh corresponding $\lambda$ values
(from Boissier et al. 2000).
}
\end{figure}

It can be seen that a correlation is always found in our models between
abundances ($Z$) and  $V_C$ or M$_B$, stronger in the case of $Z$(3 kpc) than
in the other two cases. Only in the $\lambda$=0.01 models (unrealistic for 
disks, as stressed already), $Z$(0.8 $R_d$) is not correlated with either $M_B$
or $V_C$. Except for that, the grid of our models reproduces fairly well the
data, both concerning absolute values and slopes of the correlations.

\subsection{Star formation efficiency}

The agreement of our models to such a large body of observational data 
(see Boissier and Prantzos 2000 for details) is
impressive, taking into account the small number of free parameters.
Indeed, the only really ``free'' parameter is the dependence of the
infall timescale on galactic mass, since all other ingredients
are fixed  by the calibration of the model to the Milky Way disk. 

We notice that our assumption about infall timescales leads to massive
disks forming their stars {\it earlier} than less massive ones. We also
notice that this is not due to the corresponding star formation (SF)
efficiency. As already stressed in Sec. 3.1, the efficiency of our adopted
SFR is independent of $V_C$ (for a given $\lambda$ value). This
assumption is justified by the data of Fig. 12 ({\it lower panel}).
The observed current global SF efficiency of disk galaxies, measured by the ratio
of the SFR to the mass of gas, seems to be independent of $V_C$,
in fair agreement with our models (of course, higher $\lambda$ values
correspond to lower efficiencies).

On the other hand, Fig. 12 ({\it upper panel}) shows also the
corresponding gas fraction as a function of $V_C$: lower mass disks
have proportionately more gas than massive ones. When the two
panels of Fig. 12 are compared, there is only one possible logical
conclusion: {\it massive galaxies formed their  stars earlier than their
low mass counterparts}.Thus, they transformed a larger fraction of their gas 
to stars and have today a redder stellar population and a higher metallicity
at a given radius [{\it Notice:} The underlying assumption is that the 
SF efficiency has not varied by much during the disk history; but, if
current SF efficiencies are similar for such a large range of galaxy masses
and metallicities, it is difficult to see why the situation should
have been very different in the past.]

These features are not easy to accomodate in the framework of
currently popular  hierarchical models of galaxy formation and evolution.
For instance, the observed ``redder'' colours of more massive disks are
usually interpreted in such models by invoking the presence of large
amounts of dust (e.g. Somerville and Primack 1999). 
Although there is no doubt that the more massive the galaxy
the larger is the overall amount of gas and  dust (see Fig. 8), the
effect should not be overestimated. Indeed, current evidence suggests that 
galactic disks are optically thin (e.g. Xilouris et al. 1997).
Moreover, the fact that
the Milky Way is ``redder'' than the Magellanic Clouds is certainly not
due to its larger dust content....

\begin{figure}[ht]
  \begin{center}
    \epsfig{file=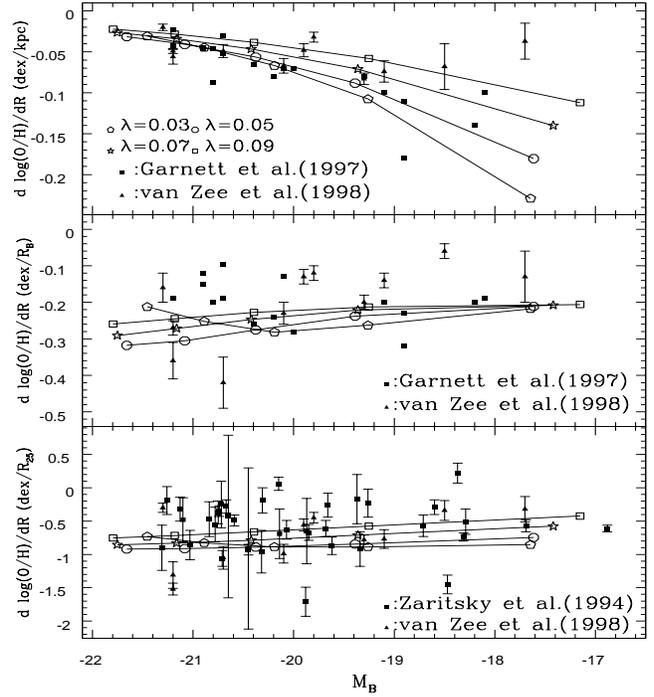, height=10cm, width=8.9cm}
  \end{center}
\caption{Oxygen abundance gradients of  our models at T=13.5 Gyr 
({\it open symbols}, corresponding
to the values of the spin parameter $\lambda$ as indicated in the upper
panel and connected with {\it solid} curves). They are plotted as a function
of the B-magnitude and compared to observations ({\it filled symbols}, with 
appropriate references given in each panel).
{\it Upper panel}: Oxygen gradient in dex/kpc;
{\it Middle panel}: Oxygen gradient in dex/R$_B$ 
($R_B$ is the disk scalelength in the $B$-band);
{\it Lower panel}: Oxygen gradient in dex/R$_{25}$
($R_{25}$ is the radius where surface brightness
is 25 mag/arcsec$^2$) (from Prantzos and Boissier 2000).}
\end{figure}

\section{Abundance gradients in spirals}

Almost all large spiral galaxies present sizeable radial abundance
gradients. This is, for instance, the case for the Milky Way disk, showing an 
oxygen abundance gradient of dlog(O/H)/dR$\sim$-0.07 dex/kpc in both its 
gaseous and stellar components. The origin of these gradients is still a matter
of debate. A radial variation of the star formation rate (SFR), or 
the existence of radial gas flows, or a combination of these processes,
can lead to abundance gradients in disks
(e.g. Henry and Worthey 1999 and references therein). 
On the other hand, the presence
of a central bar inducing large scale mixing through radial gas flows
tends to level out preexisting abundance gradients
(e.g. Dutil and Roy 1999). Despite several studies
in the 90ies (e.g. Friedli et al. 1998 and references therein) 
the relative importance of these processes has not been clarified yet.

\begin{figure}[ht]
  \begin{center}
    \epsfig{file=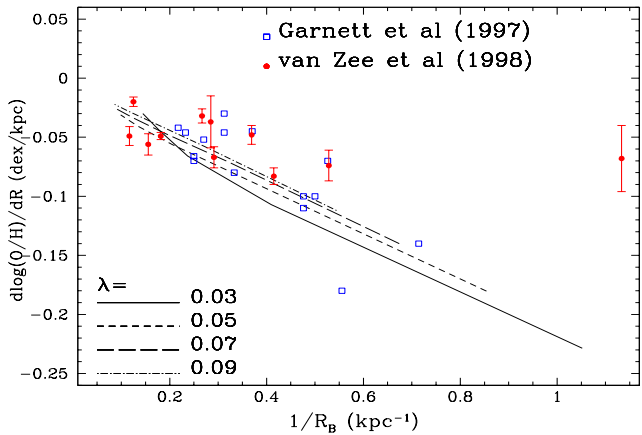, height=8.9cm, width=6cm, angle=-90}
  \end{center}
\caption{Predictions of our models for the gradient of metallicity 
(in dex/kpc) vs 1/R$_B$ and comparison to observations. 
Our model results  are parametrised by the corresponding values of the 
spin parameter $\lambda$.  A clear correlation is found between abundance 
gradient (in dex/kpc) and 1/R$_B$ (from Prantzos and Boissier 2000).}
\end{figure}

In Fig. 13, we compare our results to the observations of Garnett et al. (1997)
and van Zee et al. (1998) concerning abundance gradients in external spirals 
(i.e. observations of O/H in HII regions as a function of galactocentric distance). 

In the upper panel, abundance gradients are expressed in dex/kpc. Observations show
that luminous disks have small gradients, while as we go to low luminosity ones
the absolute value of the gradients and their dispersion increase.
This trend is fairly well reproduced by our models. Indeed,
the $\Sigma_g^{1.5}$/R factor in
the adopted SFR creates large abundance variations within short distances in
small galaxies. On the other hand, important gradients cannot
 be created in massive disks, where neighboring 
regions differ little in $\Sigma_g$.

In the middle panel of Fig. 13, the oxygen gradients are expressed in dex/R$_B$.
When expressed in this unit, the observed abundance gradients show no more any 
correlation to M$_B$, as already noticed in Garnett et al. (1997). Moreover,
a considerable dispersion is obtained for all M$_B$ values,
while the average gradient is $\sim$-0.2 dex/$R_B$.
Since the estimates of R$_B$, both in observations and in our models,
may be affected by extinction, we consider that the agreement 
of our models with the data
is quite satisfactory. Another difficulty may stem from the fact that
relatively few disks can be fit with perfect exponentials; according to
Courteau and Rix (1999) this happens for only $\sim$20\% of the disk galaxies. 

In the lower panel of Fig. 13 abundance gradients are expressed in 
dex/R$_{25}$. Again, observations show no trend with M$_B$ and a
considerable dispersion. 
Our results also show no correlation with M$_B$,
while the average model value ($\sim$-0.8 dex/R$_{25}$) is in perfect
agreement with observations. 
This is most encouraging, since $R_{25}$ is less affected by 
considerations on dust extinction or fit to an exponential
profile.

If one combines the conclusions of the upper and middle panel of Fig. 13,
namely the facts that:
a) abundance gradients (in dex/kpc) become more important and present
a larger dispersion in low luminosity disks and
b) abundance gradients in dex/R$_B$ are independent of disk luminosity,  
then she/he immediately concludes that there must be a one-to-one
correlation between abundance gradients and scalelength R$_B$.
The correlation must be relatively tight, since the 
observed dispersion ({\it middle panel} of Fig.  13) is relatively small.
Although this conclusion is an immediate consequence
of the observations in Fig. 13, it has only  been noticed
recently (Prantzos and Boissier 2000).

In Fig. 14 we plot the oxygen abundance gradients (in dex/kpc) 
as a function of  1/R$_B$ 
(which is proportional to a ``magnitude gradient'' in mag/kpc), 
 both from observations and from our  models.
It can be seen that
our models show an excellent correlation between the abundance gradients and
1/R$_B$: smaller disks have larger gradients and the results depend
little on $\lambda$. In principle, 
{\it knowing the B-band scalelength of a disk,
one should be able to determine the corresponding abundance gradient and 
vice-versa}! In practice, however, observations show a scatter around the
observed trend. 
Since large
disks ``cluster'' on the upper left corner of the diagram, we
suggest that observations of the abundance gradients in small disks would
help to establish the exact form of the correlation.
The success of our models (calibrated on the Milky Way)
in reproducing the observations points to
a ``homologuous evolution'' for galactic disks, as suggested by Garnett
et al. (1997) but does not explain the origin of the gradients.

\begin{acknowledgements}

 The results presented in this paper have been obtained in collaboration with 
S. Boissier and are part of his PhD Thesis.

\end{acknowledgements}

\end{document}